\documentclass[showpacs,preprintnumbers,twocolumn,amsmath,amssymb]{revtex4}
\usepackage{graphicx}
\usepackage{subfigure}
\usepackage{dcolumn}
\usepackage{bm}

\begin{document}

\title{Microscopic reversibility of quantum open systems}
\author{Takaaki Monnai}%
\email{monnai@suou.waseda.jp}%

\affiliation{$*$Department of Applied Physics, Osaka City University,
3-3-138 Sugimoto, Sumiyoshi-ku, Osaka 558-8585, Japan}
\begin{abstract}
The transition probability for time-dependent unitary evolution is invariant under the reversal of protocols.
In this article, we generalize the expression of microscopic reversibility to externally perturbed large quantum open systems. The time-dependent external perturbation acts on the subsystem during a transient duration, and subsequently the perturbation is switched off so that the total system would thermalize.
We concern with the transition probability for the subsystem between the initial and final eigenstates of the subsystem. In the course of time evolution, the energy is irreversibly exchanged between the subsystem and reservoir.    
The time reversed probability is given by the reversal of the protocol and the initial ensemble. Microscopic reversibility equates the time forward and reversed probabilities, and therefore appears as a thermodynamic symmetry for open quantum systems.
\end{abstract}
\maketitle
\section{Introduction}
The time reversal invariance of the equations of motion amounts to universal symmetry of fluctuation theorems and related equalities\cite{Jarzynski,Evans1,Gallavotti,Kurchan,TasakiMatsui,Maes,Gelin,Jarzynski2,Crooks,Crooks2,Crooks3,Talkner,Andrieux,Andrieux2,Monnai,Monnai2,Utsumi,Esposito,Esposito2} for the fluctuation of entropy production and particle current for mesoscopic systems, and thus plays fundamental role in the nonequilibrium statistical mechanics.
The symmetries connect the probabilities of positive and negative entropy production\cite{Crooks,Esposito3}, which are 
calculated from the time forward and reversed transition probabilities for quantum systems\cite{Kurchan,TasakiMatsui,Andrieux,Monnai,Monnai2,Utsumi}.
In the classical Markovian stochastic dynamics, the {\it conditional} probability functional of the trajectories satisfies a symmetry expressed by the probability functionals and heat.
This relation is also called microscopic reversibility\cite{Crooks}, and generalized to the quantum open systems by concerning the heat calculated from set of transitions of the reservoir states\cite{Crooks2,Crooks3}.
On the contrary, we are interested in the transition probabilities between the system states instead of the statistics for dissipative quantities, and give an expression of microscopic reversibility.
It is known that the induced absorption and emission by an external electric field are equally probable within the realm of the Fermi's Golden rule provided that the initial and final Fock states are exchanged.
Absorption process can be seen as the time reversal of the corresponding emission process.
   
In this article, we show that a symmetry similar to that of the induced absorption holds for the macroscopic time-dependent open systems where the system couples to the reservoir. Since the expression appears as a generic equation for the forward and reversed protocols, we call it as microscopic reversibility.

This article is organized as follows.
First, we describe our model and forcing protocol.
And an expression of microscopic reversibility for open systems is derived in Eq.(\ref{cyclic}).
Then the microscopic reversibility is numerically verified.
\section{Model}
Let us consider a finite system interacting with a macroscopically large reservoir at an inverse temperature $\beta$. 
The system is externally controlled by a time dependent parameter $\lambda(t)$, which is for example a spring constant for the case of a harmonic oscillator.
Therefore the total energy change of the system is caused by the external work done and the heat flow from the reservoir. 
The total Hamiltonian is
\begin{equation}
H(t)=H_s(\lambda(t))+H_r+H_{sr},
\end{equation}
where $H_s(\lambda(t))$, $H_r$, and $H_{sr}$ are the Hamiltonians of the system, the reservoir, and the interaction between them, respectively. 

Let us prepare the initial state as 
\begin{eqnarray}
\rho(0)&=&\rho_s(\lambda(0))\otimes \rho_r \; ;\nonumber \\
\rho_s(\lambda(0))&=&\frac{e^{-\beta H_s(\lambda(0))}}{Z(\lambda(0))} 
\end{eqnarray}
which is the product of the canonical ensembles of the system $\rho_s(\lambda(0))$ and of the reservoir $\rho_r$ at the same inverse temperature $\beta$.
The partition function of the system is 
\begin{equation}
Z(\lambda(0))\equiv{\rm Tr_s}e^{-\beta H_s(\lambda(0))}.
\end{equation} 
Through out this paper, we assume that the interaction energy $H_{sr}$ is small compared to the energy of the subsystem $H_s$ and the reservoir $H_r$.
This assumption is reasonable for macroscopic systems, since $H_s$ and $H_r$ are the bulk energy, while $H_{sr}$ would be proportional to the surface area.   
Note that the assumption is different from the weak coupling limit.
Namely, the weak coupling assumes that interaction is negligible, and the perturbative analysis is available.
On the other hand, in our case the interaction is not necessarily vanishing, while we require that the ratio between the interaction energy and bulk energy is negligible.
\section{Twice measurements scheme}
The twice measurements scheme consists of the initial and final observations, (i) and (ii).
\begin{itemize}
\item[(i)] At $t=0$, we measure the energy operator $H_s(\lambda(0))$, and gain an eigenenergy $E_n(0)$. The system state becomes the corresponding eigenstate $|n(0)\rangle$, 
\begin{equation}
H_s(\lambda(0))|n(0)\rangle =E_n(0)|n(0)\rangle.
\end{equation}
Subsequently the total system unitarily evolves until $t=T$ as 
\begin{equation}
\rho(T)=U(|n(0)\rangle\frac{e^{-\beta E_n(0)}}{Z(\lambda(0))}\langle n(0)|\otimes\rho_r) U^+.
\end{equation}
Here $U={\rm T}\{e^{-\frac{i}{\hbar}\int_0^T ds H(s)}\}$ is the unitary time evolution operator.
For the time evolution, we require that the external forcing is switched off well in advance $t=T$ and {\it the total density matrix would relax to an equilibrium state} at $t=T$,  
\begin{eqnarray}
U\rho(0)U^+&\cong& \rho_s(\lambda(T))\otimes \rho_r ;\nonumber \\
&&\rho_s(\lambda(T))=\frac{e^{-\beta H_s(\lambda(T))}}{Z(\lambda(T))}.\label{relaxation}
\end{eqnarray}
Here $H_s(\lambda(T))$ is the corresponding energy operator at $t=T$ and $Z(\lambda(T))$ is the partition function.
Eq.(\ref{relaxation}) comes from the equation at the level of matrix elements\cite{Lebowitz}
\begin{equation}
{\rm Tr}_r U\rho(0)U^+\cong \rho_s(\lambda(T)) \label{relaxation2}
\end{equation}
and the assumption of smallness of the interaction energy, and holds when acting on the local state of the subsystem as in Eq.(\ref{cyclic}).
\item[(ii)] At $t=T$, we measure the energy 
$H_s(\lambda(T))$ 
and obtain some eigenenergy $E_m(T)$ with the corresponding eigenvector $|m(T)\rangle$.
{\it Regarding the reservoir, we don't perform any measurements}.
The transition probability that the initial and final system states are $|n(0)\rangle$ and $|m(T)\rangle$ is then    
\begin{eqnarray}
&&P_F(|n(0)\rangle\rightarrow|m(T)\rangle) \nonumber \\
&=&{\rm Tr_r}\{\langle m(T)|U|n(0)\rangle\frac{e^{-\beta E_n(0)}}{Z(\lambda(0))}\rho_r\langle n(0)|U^+|m(T)\rangle\}.\label{eq} \nonumber \\
\end{eqnarray}
Here the Kraus operator 
\begin{equation}
A_{nm}\equiv\frac{1}{\sqrt{Z(\lambda(0))}}e^{-\frac{\beta E_n(0)}{2}}\langle m(T)|U|n(0)\rangle
\end{equation} describes the transition of the system state, and satisfies 
\begin{eqnarray} 
&&\sum_{n,m}A_{nm}^+A_{nm}=\sum_n \langle n(0)|\frac{1}{Z(\lambda(0))}e^{-\beta H_s(\lambda(0))}U^+\nonumber \\
&&\sum_m|m(T)\rangle\langle m(T)|U|n(0)\rangle=1.
\end{eqnarray}  
Note that the matrix elements such as $\langle m(T)|U|n(0)\rangle$ contain the reservoir variables. 
The second equality follows from the completeness $\sum_m |m(T)\rangle\langle m(T)|=1$, and the unitarity $U^+U=1$.

Similarly the probability of the time-reversed dynamics is calculated as well.
Firstly, let us define the reversed dynamics by reversing the time dependence of the system Hamiltonian in Eq.(2) as $H_s(\lambda(T-t))$, i.e. the reversal at time $T$. 
Also, we start with the initial state $\Theta\frac{e^{-\beta H_s(\lambda(T))}}{Z(\lambda(T))}\otimes \rho_r\Theta^{-1}$, where $\Theta$ is the anti unitary time reversal operator of the total system.
The partition function is defined as 
\begin{equation}
Z(\lambda(T))\equiv{\rm Tr_s}e^{-\beta H_s(\lambda(T))}.
\end{equation}

\item[(i-2)]We measure the energy $\Theta H_s(\lambda(T))\Theta^{-1}$ at $t=0$ and consider the case that the eigenenergy $E_m(T)$ corresponding to (i), and the system state becomes $\Theta|m(T)\rangle$.
$\Theta|m(T)\rangle$ is proportional to $|m(T)\rangle$ when the energy $E_m(T)$ does not degenerate. 
\item[(ii-2)]At the final time $t=T$, we again measure the energy $\Theta H_s(\lambda(0))\Theta^{-1}$ and obtain the eigenenergy $E_n(0)$.  
The system state is $\Theta|n(0)\rangle$.
 
Then the probability that the initial and final states are $\Theta|m(T)\rangle$ and $\Theta|n(0)\rangle$ for the reversed dynamics is
\begin{eqnarray}
&&P_R(\Theta|m(T)\rangle\rightarrow \Theta|n(0)\rangle) \nonumber \\
&=&{\rm Tr_r}\{\langle n(0)|\overleftarrow{\Theta}U^+\Theta|m(T)\rangle\frac{e^{-\beta E_m(T)}}{Z(\lambda(T))}\rho_r\langle m(T)|\overleftarrow{\Theta}U\Theta|n(0)\rangle\}. \nonumber \\
&&
\end{eqnarray}
Here $\langle n(0)|\overleftarrow{\Theta}U^+\Theta|m(T)\rangle=\langle m(T)|U|n(0)\rangle$ is the inner product of $\Theta|n(0)\rangle$ and $U^+\Theta|m(T)\rangle$. 
\end{itemize}
\section{Microscopic reversibility}
Now let us show the relation between probabilities of the forward and reversed transitions
\begin{equation}
P_F(|m(0)\rangle\rightarrow|n(T)\rangle)\cong P_R(\Theta|n(T)\rangle\rightarrow\Theta|m(0)\rangle).\label{reverse}
\end{equation}
Here the difference between the forward and reversed probabilities goes to zero in the macroscopic limit.  
For this purpose, we apply the methodology developed in Refs.\cite{Crooks2,Crooks3} and thermalization property Eq.(\ref{relaxation}).
We rewrite the forward probability as 
\begin{eqnarray}
&&P_F(|n(0)\rangle\rightarrow |m(T)\rangle)\nonumber \\
&=&{\rm Tr_r}\rho_r^{\frac{1}{2}}\langle n(0)|\frac{1}{\sqrt{Z(\lambda(0))}}e^{-\frac{1}{2}\beta H_s(\lambda(0))}U^+|m(T)\rangle\nonumber \\
&&\rho_r^{-\frac{1}{2}}\rho_r\rho_r^{-\frac{1}{2}}\langle m(T)|U\frac{1}{\sqrt{Z(\lambda(0))}}e^{-\frac{1}{2}\beta H_s(\lambda(0))}|n(0)\rangle\rho_r^{\frac{1}{2}}\nonumber \\
&=&{\rm Tr_r}\langle n(0)|U^+(\{U\rho_s(\lambda(0))^{\frac{1}{2}}\rho_r^{\frac{1}{2}}
U^+\}\rho_s(\lambda(T))^{-\frac{1}{2}}\rho_r^{-\frac{1}{2}})\nonumber \\
&&\rho_s(\lambda(T))^{\frac{1}{2}}|m(T)\rangle\rho_r\langle m(T)|\rho_s(\lambda(T))^{\frac{1}{2}}\nonumber \\
&&(\rho_s(\lambda(T))^{-\frac{1}{2}}\rho_r^{-\frac{1}{2}}\{U\rho_s(\lambda(0))^{\frac{1}{2}}  
\rho_r^{\frac{1}{2}}U^+\})U|n(0)\rangle\nonumber \\
&\cong&{\rm Tr_r}\langle n(0)|U^+\frac{1}{\sqrt{Z(\lambda(T))}}e^{-\frac{1}{2}\beta H_s(\lambda(T))}|m(T)\rangle\rho_r\nonumber \\
&&\langle m(T)|\frac{1}{\sqrt{Z(\lambda(T))}}e^{-\frac{1}{2}\beta H_s(\lambda(T))}U|n(0)\rangle\nonumber \\
&=&\frac{1}{Z(\lambda(T))}e^{-\beta E_m(T)}{\rm Tr_r}\langle n(0)|U^+|m(T)\rangle\rho_r\langle m(T)|U|n(0)\rangle \nonumber \\
&=&P_R(\Theta|m(T)\rangle\rightarrow \Theta|n(0)\rangle). \label{cyclic}
\end{eqnarray}
The first equality derives from the cyclic property of the trace. In the second equality, $U^+U=1$ is inserted. 
Also, with the use of the relaxation property of the density matrix Eq.(\ref{reverse}) which comes from Eq.(\ref{relaxation2}) and the smallness of interaction energy $H_{sr}$ with respective to the bulk energy, the contents of the curly brackets are just the inverse of the 
remaining quantities in the brackets $\rho_s(\lambda(T))^{-\frac{1}{2}}\rho_r^{-\frac{1}{2}}$.
\section{Numerical demonstration of microscopic reversibility}
In this section, we numerically show the microscopic reversibility Eq.(\ref{cyclic}).
See also the detailed numerical verification in Ref.\cite{Kawamoto} for various parameters and measurement basis.  
We consider $N=8$ site spin chain in the spatially inhomogeneous time dependent magnetic field.
Note that thermalization property has been observed in relatively small system sizes\cite{Saito,Shankar}.
In the context of quantum derivation of thermal state, the dimension of the Hilbert space, which exponentially depends on the system size plays essential role\cite{Lebowitz}.     
The Hamiltonian is given as
\begin{eqnarray}
&&H(t)=H_s(t)+H_r(t)+H_{sr} \nonumber \\
&&H_s(t)=-J_s\sum_{j=1}^{N_s-1}\sigma_j^z\sigma_{j+1}^z+\alpha_s\sum_{j=1}^{N_s-1}\sigma_j^x+h(t)\sum_{j=1}^{N_s-1}\sigma_j^z \nonumber \\
&&H_r=-J_r\sum_{j=N_s+1}^{N}\sigma_j^z\sigma_{j+1}^z+\alpha_r\sum_{j=N_s+1}^{N}\sigma_j^x+\gamma_r\sum_{j=N_s+1}^N\sigma_j^z \nonumber \\
&&H_{sr}=-J_{sr}\sigma_{N_s}^z\sigma_{N_s+1}^z+\alpha_{sr}\sigma_{N_s}^x+\gamma_{sr}\sigma_{N_s}^z,
\end{eqnarray}
where $\sigma_j^i$ is the $i$ component of the Pauli matrix.
The sites $j=1,2(N_s=2)$ and $3<j\leq N$ are regarded as the subsystem and reservoir.
The external magnetic field $h(t)=\frac{B}{2}-\frac{B}{2}\tanh(\mu(t-\tau))$ satisfies $h(0)\cong B$ and switched off after $t=\tau$.
The initial state is prepared as 
\begin{eqnarray}
&&\rho(0)=\rho_s(0)\rho_r\; ; \nonumber \\
&&\rho_s(0)=\frac{1}{Z_s(0)}e^{-\beta H_s(0)} \nonumber \\
&&\rho_r=\frac{1}{Z_r}e^{-\beta H_r}.
\end{eqnarray}
We calculated the forward and reversed transition probabilities in Fig.1.
The parameters are chosen as follows. The exchange interactions at each site are $J_s=J_r=J_{sr}=1$, the $x$ component of the magnetic field is expressed by the parameters $\alpha_s=\alpha_r=1$, $\alpha_{sr}=0.2$, and similarly the $z$ component of the magnetic field is given by $h(t)$ with $B=3$, $\mu=5$, $\gamma_r=1$, and $\gamma_{sr}=0.2$.
The inverse temperature is $\beta=0.01$ and the switching time is $\tau=5$.
 
The unitary time evolution $U={\rm T}\{e^{-i\int_0^T H(t)dt}\}$ is discretized as $e^{-i\Delta t H(N\Delta t)}e^{-i\Delta t H((N-1)\Delta t)}\cdot\cdot\cdot e^{-i\Delta t H(\Delta t)}e^{-i\Delta t H(0)}$ with the time step $\Delta t=0.05$ which is much shorter than the time scale of   external perturbation.
Initially, we measure the eigenstate of $H_s(0)$ corresponding to the largest eigenvalue.
The measurement basis at $t=T$ is chosen as the eigenstate of $H_s(T)$ which corresponds to the second largest eigenvalue. Note that there are no crossing of time dependent energy levels of the subsystem, and Eq.(\ref{reverse}) holds also for the transitions to other three eigenstates.   
It is remarkable that the time dependence of the forward and reversed transition probabilities are similar as a function of time even after the quench at $t\cong \tau$. 
\begin{figure}
\center{
\includegraphics[scale=0.8]{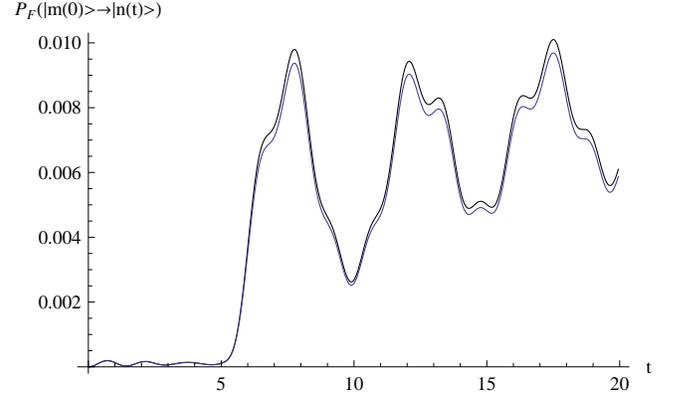}
}
\caption{The time dependence of the forward and reversed probabilities $P_F(|m(0)\rangle\rightarrow |n(t)\rangle)$(blue line) and $P_R(\Theta|n(t)\rangle\rightarrow \Theta|m(0)\rangle)$(black line) is shown where the initial and states are the eigenstates with the first to second largest eigenvalues.}
\end{figure}
\section{Summary}
We have derived an expression of microscopic reversibility for macroscopic quantum open systems.
The microscopic reversibility is trivial for the transition probability of unitary time evolution for the total system. Similar reversibility is well-known for the transition probabilities of induced absorption and emission under the influence of an electric field in equilibrium. The microscopic reversibility is regarded as a generalization to the case of generic time-dependent perturbation. Therefore, it is a symmetry holds in generic macroscopic quantum open systems.    
The microscopic reversibility is numerically verified for a spin chain with time-dependent perturbation. 
In the context of the quantum generalization of the reversibility for the classical conditional probability functional, Ref.\cite{Crooks2} derives another symmetry for the quantum trajectories.      
The main difference from Ref.\cite{Crooks2} is the quantity which we measure, number of measurements, and the definition of the reversed process.
Here we pursue and measure the system states.   
This is a useful property of the present scheme, since the system variables are expected to be much easier to measure with sufficient accuracy compared to those of the spatially extended large reservoir.

\section{Acknowledgment}
T.M. is grateful to Professor S.Tasaki for his encouragement and Professor A.Sugita for fruitful discussions.
This work is supported by the JSPS research program under the Grant 22$\cdot$7744.

\end{document}